\documentclass[
 a4paper,
 superscriptaddress,
 amsmath,amssymb, twocolumn,
 prr,
 longbibliography,
 floatfix,
 notitlepage,
 showpacs,
 noeprint
]{revtex4-2}

\usepackage{amsmath}
\usepackage{amsfonts}
\usepackage{amssymb}
\usepackage{xcolor}
\usepackage{braket}

\usepackage[colorlinks=true,citecolor=blue]{hyperref}
\usepackage[normalem]{ulem}

\usepackage{graphicx}

\begin{document}
\title{Preparation of conditionally-squeezed states in qubit-oscillator systems}

\author{Marius K. Hope}
\affiliation{Norwegian National Security Authority, Norway}
\affiliation{Department of Science and Industry Systems, University of
  South-Eastern Norway, PO Box 235, Kongsberg, Norway}

\author{Jonas Lidal}
\affiliation{Norwegian National Security Authority, Norway}

\author{Francesco Massel}
\affiliation{Department of Science and Industry Systems, University of
  South-Eastern Norway, PO Box 235, Kongsberg, Norway}

\begin{abstract}
Inspired by recent advances in the manipulation of superconducting circuits
coupled to mechanical modes in the quantum regime, we propose a protocol for
generating superpositions of orthogonally squeezed states in a quantum harmonic
oscillator.  The protocol relies on a quadratic coupling between the oscillator
and a qubit, and is conceptually similar to methods used for preparing cat states in qubit-oscillator systems. We numerically evaluate the robustness of the state-preparation scheme in the presence of decoherence, considering environmental coupling for both the harmonic oscillator and the qubit. As a potential application, we introduce a quantum error-correcting code based on conditionally-squeezed states and analyze its error-mitigation properties.
\end{abstract}

\maketitle

\paragraph{Introduction.}

Quantum technologies rely on the ability to design and manipulate systems that are fundamentally quantum mechanical~\cite{dowlingQuantumTechnologySecond2003}. By harnessing quantum effects such as superposition and entanglement, these technologies are expected to profoundly impact various fields, including sensing~\cite{degenQuantumSensing2017}, information processing and communication~\cite{nielsenQuantumComputationQuantum2010}, and the simulation of quantum systems~\cite{johnsonWhatQuantumSimulator2014}.

A broad class of quantum systems relevant to quantum technologies can be
described by a prototypical model consisting of one or more two-level systems
coupled to one or more bosonic modes, under the influence of external drives. In
its versatility, this model encompasses a range of physical scenarios, from
matter-radiation interaction~\cite{wallsQuantumOptics2008}, to, more generally,
cavity QED experiments~\cite{waltherCavityQuantumElectrodynamics2006}, and to superconducting circuit
QED~\cite{blaisCavityQuantumElectrodynamics2004} setups.

In the latter context, researchers have focused on bosonic quantum error correction (bQEC) in superconducting circuits~\cite{caiBosonicQuantumError2021}. A notable recent example of bQEC is the work of Sivak et al.\cite{sivakRealtimeQuantumError2023}, which demonstrated that quantum error correction can surpass the so-called break-even point. This achievement hinges on the ability to generate specific states — GKP states — that form the foundation of the error-correction protocol. This exemplifies how the efficient preparation of high-fidelity quantum states is of both practical and fundamental significance.

While a large body of work has focused on the bosonic modes implemented as microwave
cavity modes, considerable research has also focused on coupling optical/microwave cavities 
and superconducting qubits to mechanical modes~\cite{oconnellQuantumGroundState2010,teufelSidebandCoolingMicromechanical2011,wollmanQuantumSqueezingMotion2015,pirkkalainenSqueezingQuantumNoise2015a,riedingerRemoteQuantumEntanglement2018,ockeloen-korppiStabilizedEntanglementMassive2018,chuCreationControlMultiphonon2018a,bildSchrodingerCatStates2023a}.
Depending on the mechanical mode’s characteristic frequency, researchers
distinguish between quantum optomechanics~\cite{aspelmeyerCavityOptomechanics2014} and
quantum acoustics~\cite{chuQuantumAcousticsSuperconducting2017a}. In optomechanics, early breakthroughs
included sideband cooling of mechanical modes to their quantum ground
state~\cite{teufelSidebandCoolingMicromechanical2011}, as well as the
squeezing~\cite{wollmanQuantumSqueezingMotion2015,pirkkalainenSqueezingQuantumNoise2015a} and
entanglement~\cite{riedingerRemoteQuantumEntanglement2018,ockeloen-korppiStabilizedEntanglementMassive2018} of
mechanical modes. More recently, experiments have succeeded in preparing Fock
\cite{chuCreationControlMultiphonon2018a}, and cat states~\cite{bildSchrodingerCatStates2023a} in high-overtone bulk acoustic resonators (HBARs).

From a theoretical perspective, early work~\cite{lawArbitraryControlQuantum1996a,krastanovUniversalControlOscillator2015} established the possibility of preparing any quantum state of a bosonic mode coupled to a qubit, under slightly different control conditions. Specifically, Law and Eberly~\cite{lawArbitraryControlQuantum1996a} demonstrated that for a resonantly coupled oscillator-qubit system, arbitrary oscillator states could be prepared by controlling the coupling strength and applying a qubit drive. Their protocol was later used to generate arbitrary states in a circuit QED architecture~\cite{hofheinzSynthesizingArbitraryQuantum2009}. Similarly, Krastanov et al.~\cite{krastanovUniversalControlOscillator2015} proposed a protocol enabling universal control of an oscillator via dispersive coupling to a qubit, which was subsequently used to prepare large cat states in a superconducting cavity~\cite{vlastakisDeterministicallyEncodingQuantum2013}. Other protocols target specific quantum states, such as the preparation of cat states via conditional displacement and qubit measurement~\cite{liaoGenerationMacroscopicSchrodingercat2016} or transient disentanglement of a qubit and an oscillator~\cite{buzekSchrodingercatStatesResonant1992,bildSchrodingerCatStates2023a}.  Beyond deterministic approaches, there is growing interest in leveraging machine learning (ML) techniques for quantum-state preparation and control~\cite{krennArtificialIntelligenceMachine2023}. Early results have demonstrated the preparation of Schrödinger cat states in a microwave circuit using neural networks~\cite{hutinPreparingSchrodingerCat2025}.

In this paper, we present a deterministic protocol for generating squeezed
states with orthogonal squeezing axes, as well as their superpositions, in a
system consisting of a bosonic mode coupled quadratically to a two-level system.
The superposition states can be viewed as the squeezing analogue of a cat state.
Moreover, we show that these states, in principle, can be applied to quantum
error correction (QEC), leading to a novel encoding scheme within the framework
of bosonic rotation codes~\cite{grimsmoQuantumComputingRotationSymmetric2020}.
This protocol is particularly relevant for qubit-oscillator systems exhibiting
strong quadratic coupling, as recently demonstrated for qubits coupled to
mechanical
oscillators~\cite{maNonclassicalEnergySqueezing2021a,manninenHybridOptomechanicalSuperconducting2024}.
Finally, we discuss the impact of environmental coupling and possible
experimental realizations.

\paragraph{The System.}
We consider a harmonic oscillator with a strong quadratic coupling to an externally driven qubit. The system is described by the Hamiltonian ($\hbar = 1$)
\begin{equation}\label{eq:hamiltonian_second_order_term}
    H(t) = \frac{\omega_q}{2}\sigma_z + \omega_m b^\dagger b + g(b + b^\dagger)^2 \sigma_z + H_d(t),
\end{equation}
where $\omega_q$ ($\omega_m$) is the qubit (oscillator) energy, $g$ is the coupling strength, $b^\dagger$ ($b$) is the bosonic raising (lowering) operator, and $\sigma_z$ is the Pauli $z$ operator. The qubit drive is described by the term $H_d(t)$.

Although our treatment is, in principle, device independent, we keep in mind
that a strong quadratic coupling in the form of
Eq.~(\ref{eq:hamiltonian_second_order_term}) was recently reported in an
electromechanical setup, coupling a Cooper-pair box to a mechanical
oscillator~\cite{maNonclassicalEnergySqueezing2021a} and
proposed~\cite{manninenHybridOptomechanicalSuperconducting2024} for a nanopillar
coupled to a superconducting circuit.

Our goal is to use the qubit drive to generate strong and controllable squeezing
of the oscillator. It is possible to show~\cite{supplemental} that this can be
achieved by a harmonic drive of the qubit  at $\omega_d =\omega_m$ in a frame rotating at $\omega_{q}$.  In the laboratory frame, the drive is of the form
\begin{equation}\label{eq:external-drive}
    H_d(t) = A \text{cos}(\omega_d t) [\text{cos}(\omega_q t) \sigma_x + \text{sin}(\omega_q t) \sigma_y],
\end{equation}
where the amplitude $A$ and frequency $\omega_d$ are assumed to be freely
adjustable parameters. In the lab frame, the relevant terms of
Eq.~(\ref{eq:external-drive}) oscillate at $\omega_q \pm \omega_d$ and therefore
can be construed as inducing sideband transitions between the qubit and the oscillator for $\omega_d \simeq \omega_m$.

Constraints and conditions for optimal squeezing are determined by the form of
the qubit-mechanics coupling Hamiltonian in the interaction picture~\cite{supplemental}. For $\omega_d \simeq \omega_m$ and $\omega_m \gg g$, we can perform a rotating-wave approximation (RWA) and obtain  
\begin{equation}\label{eq:hamiltonian_rwa}
    \tilde{H}_{\text{RWA}} =  g J_0(\bar{A})(2b^\dagger b +1) \sigma_z + gJ_2(\bar{A})(b^2 + b^{\dagger 2})\sigma_z,
\end{equation}
where $\bar{A} \equiv 2A/\omega_d$ and $J_n$ is the Bessel function of the first kind and $n$'th order. The two terms in $\tilde{H}_{\text{RWA}}$ can be though of as a qubit splitting dependent on the occupation number of the oscillator and a conditional squeezing term where the squeezing parameter depends on the state of the qubit. When the amplitude is such that $J_0(\bar{A}) = 0$, the first term vanishes and $\tilde{H}_{\text{RWA}}$ reduces to a simple conditional squeezing Hamiltonian
\begin{equation}\label{eq:hamiltonian_cs}
    H_{\text{cs}} = g_{\text{cs}} (b^2 + b^{\dagger 2})\sigma_z,
\end{equation}
where $g_{\text{cs}} \equiv g J_2(\bar{A})$. Choosing the strongest possible squeezing results in the condition $\bar{A} \approx 2.405$, i.e. the first root of $J_0$.

\paragraph{Preparation of squeezed states.}
We now consider the dynamics generated by $H_{\text{cs}}$. If the qubit is in the +1 eigenstate of $\sigma_z$, it is straightforward to show that the time-evolution operator for the oscillator is equivalent to the single mode squeeze operator $S(\xi) = \text{exp}[\frac{1}{2}(\xi^* b^2 - \xi b^{\dagger 2})]$ with squeezing parameter $\xi \equiv 2ig_{\text{cs}}t$. Similarly, when the qubit is in the -1 eigenstate, the time-evolution of the oscillator is governed by $S(-\xi)$. In the phase space of the oscillator these two cases correspond to squeezing along orthogonal axes.

\begin{figure}
    \centering
\includegraphics[width=0.9\columnwidth]{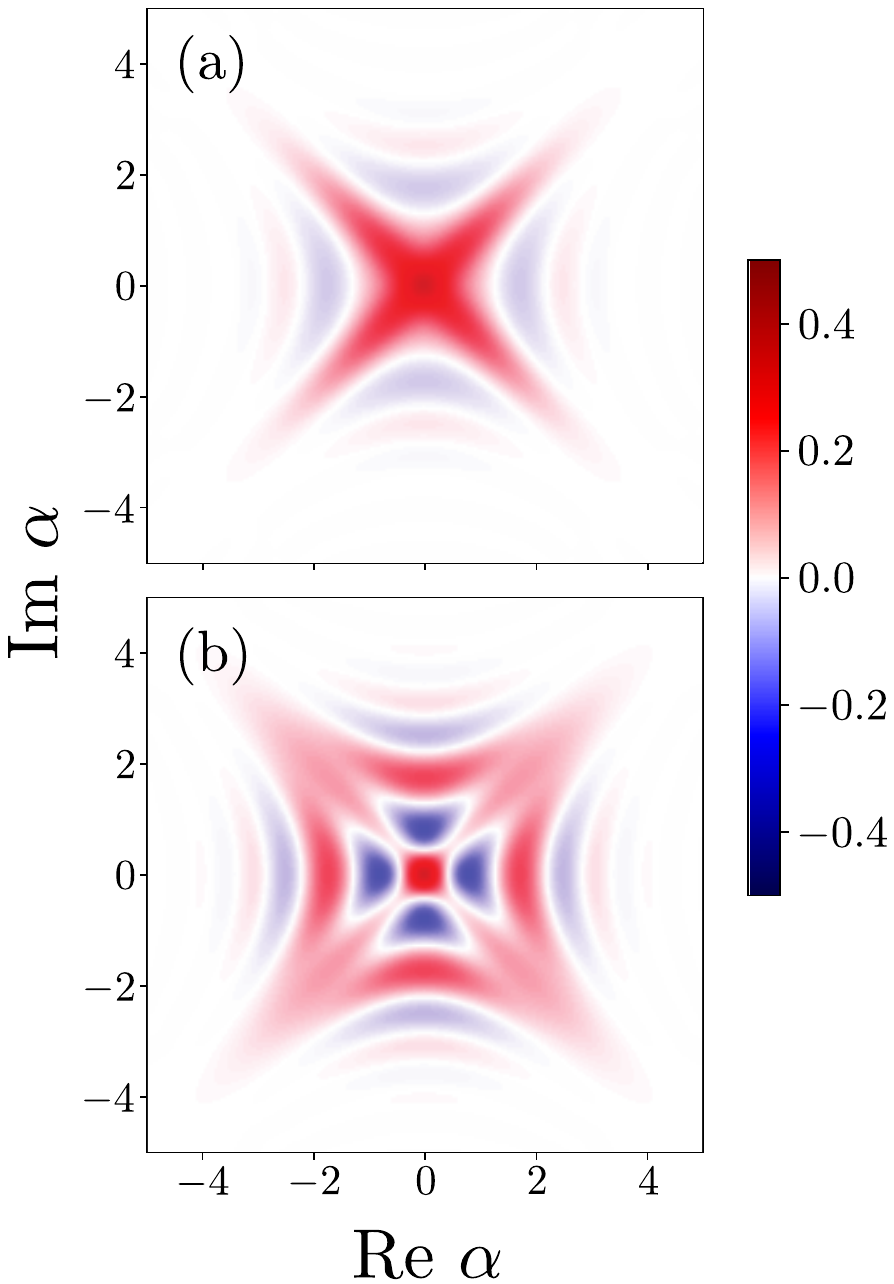}
    \caption{Wigner functions $W(\alpha)$ of (a) the symmetric and (b) the antisymmetric superposition of orthogonally squeezed states, with squeezing parameter $\xi = i$. The states can be prepared by a conditional squeezing protocol: an initial state $\ket{0}\otimes [\ket{e}+\ket{g}]$ evolves according to $H_{\text{cs}}$, followed by a projective measurement of the qubit in the $\sigma_x$ basis. The symmetric (antisymmetric) superposition state is obtained in the oscillator when the qubit is measured to be in the state $\ket{e}+ \ket{g}(\ket{e} - \ket{g})$. More generally, if the qubit is initially prepared in one of the $\sigma_x$ eigenstates, the symmetric (antisymmetric) superposition is obtained in the oscillator when the qubit is later measured to be in the same (different) state.}
    \label{fig:logical-states}
\end{figure}

Building on these results, we now describe a protocol to generate symmetric and antisymmetric superpositions of orthogonally squeezed states, similar to how even and odd cat states can be generated by conditional displacement in qubit-oscillator systems with linear coupling \cite{liaoGenerationMacroscopicSchrodingercat2016}. We consider an initial product state with a ground state oscillator and a qubit in one of the $\sigma_x$ eigenstates. Without loss of generality, we take the initial state to be $\ket{\psi(0)} \propto \ket{0} \otimes [\ket{e} + \ket{g}]$ (where $\ket{e}$ and $\ket{g}$ are the $\sigma_z$ eigenstates). When the system evolves according to $H_{\text{cs}}$, the qubit and oscillator become entangled, and the state after time $t$ can be written as $\ket{\psi(t)} \propto [S(\xi) \ket{0}\otimes \ket{e} + S(-\xi)\ket{0} \otimes\ket{g}]$. By expressing $\ket{e}$ and $\ket{g}$ in terms of the $\sigma_x$ eigenstates, one can verify that a projective measurement of the qubit in the $\sigma_x$ basis leaves the oscillator in either the symmetric or antisymmetric superposition of the squeezed vacuum states, depending on the measurement outcome. Fig.~\ref{fig:logical-states} depicts the Wigner functions of the symmetric and antisymmetric superposition states with a squeezing parameter $\xi = i$, corresponding to a time $t = 1/(2g_{\text{cs}})$ under the evolution generated by $H_{\text{cs}}$. 

\begin{figure}
    \centering
\includegraphics[width=0.95\columnwidth]{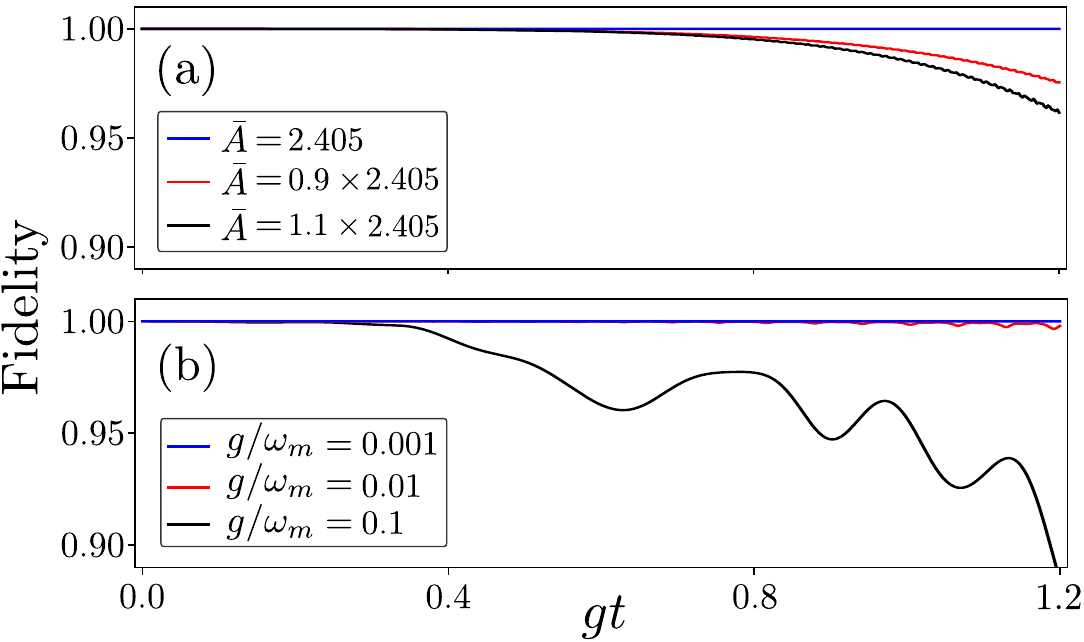}
    \caption{Sensitivity of the state preparation protocol to the conditions on (a) driving amplitude $\bar{A}$ and (b) relative coupling strength $g/\omega_m$. The fidelity is taken between the symmetric superposition state generated by $H_{\text{cs}}$ and the state generated by the full Hamiltonian in Eq.~(\ref{eq:hamiltonian_second_order_term}). The end time is chosen such that the magnitude of the squeezing parameter $\xi$ approaches 1 at the end.}
    \label{fig:fidelity_rwa_and_amplitude}
\end{figure}

To analyze the sensitivity of the state preparation to the RWA and the amplitude condition in Eq.~(\ref{eq:hamiltonian_cs}), we plot in Fig.~\ref{fig:fidelity_rwa_and_amplitude} the fidelity between the symmetric superposition generated by $H_{\text{cs}}$ and the state generated by the full Hamiltonian in Eq.~(\ref{eq:hamiltonian_second_order_term}) for different coupling strengths and drive amplitudes.

\paragraph{Application as a quantum code.}
As previously mentioned, the states in Fig.~\ref{fig:logical-states} are, in some sense, the squeezing analogue of the even and odd cat states. It is therefore tempting to define a QEC code analogous to the two-component cat code \cite{cochraneMacroscopicallyDistinctQuantumsuperposition1999} in the following way

\begin{subequations}\label{eq:logical-states}
\begin{align}
 \ket{0_L}&= \frac{1}{\sqrt{\mathcal{N}}_+}[S(\xi) + S(-\xi)]\ket{0},\\
 \ket{1_L} &= \frac{1}{\sqrt{\mathcal{N}}_-}[S(\xi) - S(-\xi)]\ket{0},
 \end{align}
\end{subequations}
where $S(\xi)\ket{0} = \frac{1}{\text{cosh} r}\sum_{n=0}^{\infty} (-1)^n \frac{\sqrt{(2n)!}}{2^n n! } e^{in \varphi} \text{tanh}^n r \ket{2n}$ is the squeezed vacuum state with complex
squeezing parameter $\xi = r e^{i\varphi}$. The normalization constants are given by $\mathcal{N}_{\pm} = 2[1\pm \text{cosh}^{-1/2}(2r)]$.

Due to the symmetry of the logical states, it is easily verified that $\ket{0}_L (\ket{1}_L)$ only contains terms of the form $\ket{4n} (\ket{4n + 2})$ in the Fock basis. In fact, this is a general property of bosonic rotation codes with 2-fold symmetry~\cite{grimsmoQuantumComputingRotationSymmetric2020}. By definition, a code has an $N$-fold symmetry if a phase space rotation of $2\pi/N$ acts as the identity operator, while a rotation by $\pi/N$ acts as a logical Pauli $z$ operator. 

Although the squeezed vacuum code in Eq.~(\ref{eq:logical-states}) was motivated
by the conceptual similarity with the two-component cat code, its symmetry in
phase space is more reminiscent of a four-component code. Indeed, it can be
shown that the code is equivalent to a four-component squeezed cat code with
zero displacement~\cite{grimsmoQuantumComputingRotationSymmetric2020}. In the
limit of large squeezing, the phase properties of a squeezed vacuum state
approaches that of a superposition of two phase states with opposite
phases~\cite{vaccaroPhaseFluctuationsSqueezing1992}. Following the definition in
Ref.~\cite{grimsmoQuantumComputingRotationSymmetric2020}, the squeezed vacuum
code is therefore an instance of the approximate number-phase codes, approaching
the ideal code in the limit of infinite squeezing. The latter, in the $N$-fold
symmetric case, can in principle detect a number shift smaller than $N$ and a phase rotation smaller than $\pi/N$~\cite{grimsmoQuantumComputingRotationSymmetric2020}. 

\begin{figure}
    \centering
\includegraphics[width=0.85\linewidth]{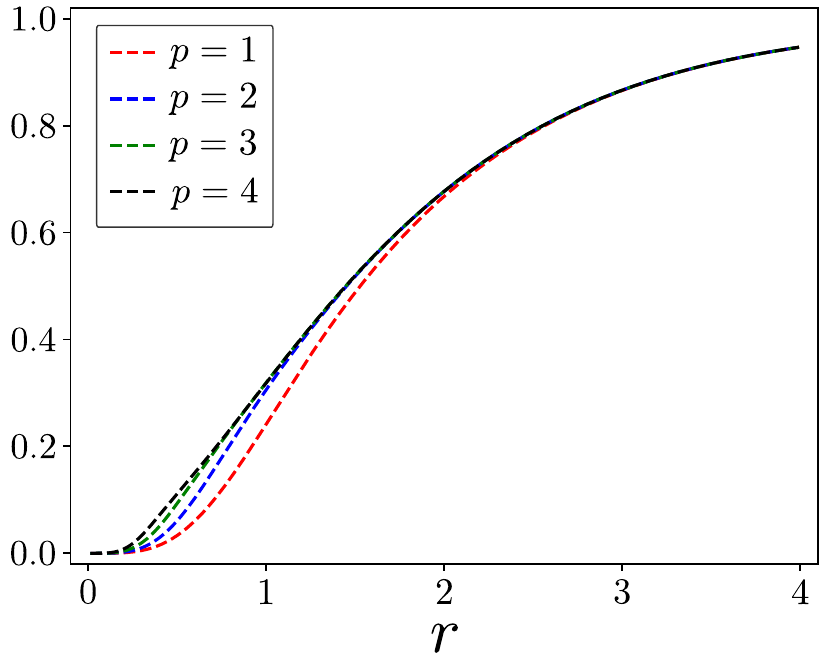}
     \caption{Ratio of the expectation values $\langle \left(b^\dagger b\right)^p\rangle$ in the two logical states of the squeezed vacuum code, i.e. $\bra{0_L}  \left(b^\dagger b\right)^p\ket{0_L}/\bra{1_L}\left(b^\dagger b\right)^p\ket{1_L}$, for $p\in \{1, 2, 3, 4\}$
    as a function of the squeezing strength $r$.}
    \label{fig:expect-value-convergence}
\end{figure}

In our setup,  we use the standard Knill-Laflamme (KL) conditions~\cite{knillTheoryQuantumErrorcorrecting1997} to evaluate the ability of the code in Eq.~(\ref{eq:logical-states}) to correct for boson loss and dephasing. Since the logical states have a distance 2 in Fock space, the code can at most detect a single boson loss event. To the leading order in the decoherence rates, the single loss and dephasing channel correspond to the error set $\mathcal{E} = \{I, b, b^\dagger b, (b^\dagger b)^2\}$, where $I$ is the identity operator~\cite{grimsmoQuantumComputingRotationSymmetric2020,schlegelQuantumErrorCorrection2022}. For a code to be able to correct the set $\mathcal{E}$, the KL conditions require that the logical states remain orthogonal after the action of any operator in $\mathcal{E}$ and that the expectation value $\langle (b^\dagger b)^p\rangle$ with $p \in \{1, 2, 3, 4\}$ is the same in both logical states. Due to the separation between the logical states in Fock space, the orthogonality condition is identically satisfied for all the operators in $\mathcal{E}$. For the same reason the expectation values are never exactly equal. However, the logical states become indistinguishable in the limit of large squeezing, analogous to the cat codes where the KL conditions are approximately satisfied for large displacements~\cite{cochraneMacroscopicallyDistinctQuantumsuperposition1999}. In Fig.~\ref{fig:expect-value-convergence} we have plotted the ratio of the expectation values in the different logical states as a function of squeezing strength.

\paragraph{Open system dynamics.}
We previously showed that we can generate the logical states in Eq.~(\ref{eq:logical-states}) with a high fidelity given certain restrictions on the coupling strength and qubit drive (Fig.~\ref{fig:fidelity_rwa_and_amplitude}). We now assume an optimal drive and evaluate how the protocol performs in an open system where dissipation effects degrade the fidelity of the prepared states. Our objective is to estimate the experimental parameter requirements to observe the superposition states shown in Fig.~\ref{fig:logical-states}.

\begin{figure}
    \centering
\includegraphics[width=0.95\linewidth]{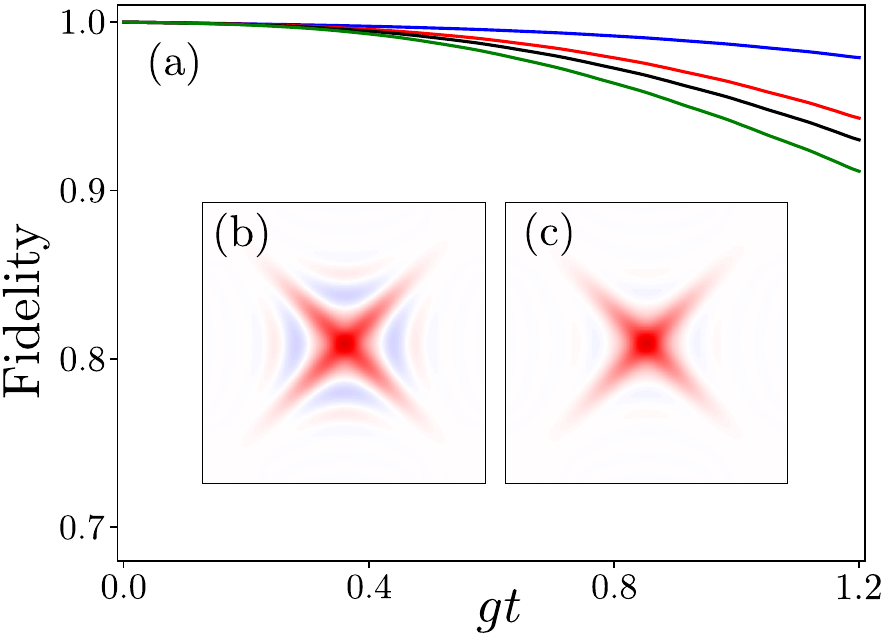}
    \caption{(a) Fidelity of the symmetric superposition state prepared in an open system with relative qubit decoherence rates $\gamma_1/g, \gamma_\phi/g = 0.1, 0.1$ (blue); $1.0, 0.1$ (red); $0.1, 1.0$ (black); $1.0, 1.0$ (green). In all cases the mechanical thermal occupation is $n_{m, th} = 1$ and relative damping rate $\gamma_m/g = 0.01$. (b) and (c) depicts the Wigner function at the end time $g t = 1.2$ for the blue and green curve respectively. }
    \label{fig:open-system-fidelity}
\end{figure}

We analyze the open system dynamics by numerically solving the master equation with the relevant jump operators~\cite{supplemental}. In principle, the state preparation is sensitive to both decay and noise in the oscillator, as well as decoherence of the qubit. Loosely speaking, the bosonic decay rate sets a limit to the level of squeezing that can be obtained in the oscillator, while the qubit decoherence limits the coherence of the bosonic superposition state. The latter is most easily seen by the weakening of the interference fringes in the Wigner plots in Fig.~\ref{fig:open-system-fidelity}~(b)-(c).

As reference we consider a mechanical oscillator with frequency $\omega_m = 1$ GHz and a superconducting qubit with energy $\omega_q = 20$ GHz coupled quadratically with a coupling strength $g = 100$ kHz. A similar setup has recently been reported in Ref.~\cite{maNonclassicalEnergySqueezing2021a} and proposed in Ref.~\cite{manninenHybridOptomechanicalSuperconducting2024}. One of the benefits of using a mechanical element as the bosonic mode, is a low intrinsic decay rate. Following Refs.~\cite{maNonclassicalEnergySqueezing2021a, manninenHybridOptomechanicalSuperconducting2024}, we will assume $\gamma_m = 1$ kHz. When the temperature is 10 mK, the thermal occupation for the oscillator and qubit baths are $n_{m, th}\approx 1$ and $n_{q, th} \approx 10^{-7}$ respectively, and the primary limitation to the state preparation is the decoherence of the qubit. In Fig.~\ref{fig:open-system-fidelity}~(a) we plot the fidelity over time of a symmetric superposition state prepared with the mentioned parameters and qubit decay ($\gamma_1$) and dephasing ($\gamma_\phi$) rates in the range $10-100$ kHz. The end time corresponds to a squeezing strength $r \approx 1$ in the closed system, and we see that successful state preparation is roughly bounded by the requirement $\gamma_1, \gamma_\phi \lesssim g = 100$ kHz. This is well within what can be achieved with superconducting qubits~\cite{burnettDecoherenceBenchmarkingSuperconducting2019,wangPracticalQuantumComputers2022} and around the values reported for superconducting qubits coupled to mechanical oscillators~\cite{maNonclassicalEnergySqueezing2021a,martiQuantumSqueezingNonlinear2024}.

\paragraph{Conclusions.}
We have shown how symmetric and antisymmetric superpositions of orthogonally squeezed vacuum states can be prepared in qubit-oscillator systems with strong quadratic coupling. The protocol uses a conditional squeezing mechanism and requires careful tuning of both frequency and amplitude of an external qubit drive. As a potential application, we proposed a bosonic code based on the prepared states and showed that the code satisfies the Knill-Laflamme conditions for single boson loss and phase errors in the limit of large squeezing.

\paragraph{Acknowledgements.}
The numerical simulations were performed using the QuantumOptics.jl numerical framework~\cite{kramerQuantumOpticsjlJuliaFramework2018}. The authors thank Juuso Mannien, Jesper Lind-Olsen and Tron Omland for fruitful discussions. FM acknowledges financial support from the Research Council of Norway (Grant No. 333937) through participation in the QuantERA ERA-NET Cofund in Quantum Technologies.

\bibliography{CondSqueeze}

\end{document}


\title{Preparation of conditionally-squeezed states in qubit-oscillator systems -- \\ SUPPLEMENTAL MATERIAL }

\author{Marius K. Hope}

\affiliation{Norwegian National Security Authority, Norway}
\affiliation{Department of Science and Industry Systems, University of
  South-Eastern Norway, PO Box 235, Kongsberg, Norway}

\author{Jonas Lidal}
\affiliation{Norwegian National Security Authority, Norway}

\author{Francesco Massel}
\affiliation{Department of Science and Industry Systems, University of
  South-Eastern Norway, PO Box 235, Kongsberg, Norway}
\maketitle

\section{Derivation of the RWA Hamiltonian}
We present here a detailed derivation of the approximate Hamiltonian $\tilde{H}_{\text{RWA}}$ from Eq.~(3) in the main text. We start with the full Hamiltonian 
\begin{equation}
     H = \frac{\omega_q}{2}\sigma_z + \omega_m b^\dagger b + g(b + b^\dagger)^2\sigma_z + H_d,
\end{equation}
where the driving term is given by 
\begin{equation}
    H_d = A\text{cos}(\omega_d t) [\text{cos}(\omega_q t)\sigma_x + \text{sin}(\omega_q t)\sigma_y].
\end{equation}
We go to a rotating frame with the unitary transformation 
\begin{equation}
    V = V_2 V_1 = \text{exp}\Big[i \frac{A}{\omega_d} \text{sin}(\omega_d t)\sigma_x\Big] \times \text{exp}\Big[i\Big(\frac{\omega_q}{2}\sigma_z + \omega_m b^\dagger b\Big) t \Big],
\end{equation}
in which the Hamiltonian is given by $\tilde{H} = V H V^\dagger + i \frac{\partial V}{\partial t} V^\dagger$. As an intermediate step, we first calculate the transformation generated by $V_1$ 
\begin{equation}\label{eq:hamiltonian_interaction_frame}
\begin{split}
    H_1 =& V_1 H V_1^\dagger + i \frac{\partial V_1}{\partial t} V_1^\dagger \\
    =& g(b^2 e^{-2i\omega_m t} + b^{\dagger 2} e^{2i\omega_m t} + 2b^\dagger b + 1 )\sigma_z \\
    &+ A\text{cos}(\omega_d t) \Big[ \text{cos}(\omega_q t) [\text{cos}(\omega_q t) \sigma_x - \text{sin}(\omega_q t)\sigma_y]
    + \text{sin}(\omega_q t)[\text{cos}(\omega_q t)\sigma_y + \text{sin}(\omega_q t) \sigma_x]\Big],
\end{split}
\end{equation}
where the last term simply reduces to $A \text{cos}(\omega_d t) \sigma_x$. The transformation generated by $V_2$ then gives the full Hamiltonian in the rotating frame
\begin{equation}
    \tilde{H} = V_2 H_1 V_2^\dagger +  i \frac{\partial V_2}{\partial t} V_2^\dagger = g(b^2 e^{-2i\omega_m t} + b^{\dagger 2} e^{2i\omega_m t} + 2b^\dagger b + 1)\Big(\text{cos}\Big[\frac{2A}{\omega_d}\text{sin}(\omega_d t)\Big] \sigma_z + \text{sin}\Big[\frac{2A}{\omega_d} \text{sin}(\omega_d t) \Big]\sigma_y \Big).
\end{equation}
Using the Jacobi-Anger expansion
\begin{equation}\label{eq:jacobi-anger}
    e^{i \chi \text{sin}(\tau)} = \sum_{n=-\infty}^{\infty} J_n(\chi) e^{in\tau},
\end{equation}
where $J_n$ is the Bessel function of the first kind and $n$'th order, together with the identity $J_{-n} = (-1)^n J_n$, we can write
\begin{equation}\label{eq:jacobi-expansion-cos-sin}
\begin{split}
    \text{cos}\Big[\frac{2A}{\omega_d} \text{sin}(\omega_d t) \Big] &= J_0\Big(\frac{2A}{\omega_d}\Big) + 2 \sum_{n = 1}^{\infty} J_{2n}\Big(\frac{2A}{\omega_d}\Big) \text{cos}(2n\omega_d t), \\
    \text{sin}\Big[\frac{2A}{\omega_d} \text{sin}(\omega_d t) \Big] &= 2 \sum_{n=1}^{\infty} J_{2n-1}\Big( \frac{2A}{\omega_d}\Big) \text{sin}([2n-1]\omega_d t).
\end{split}
\end{equation}
The full Hamiltonian in the rotating frame can thus be written as 
\begin{equation}
\begin{split}
    \tilde{H} =& g(b^2 e^{-2i\omega_m t} + b^{\dagger 2} e^{2i\omega_m t} + 2b^\dagger b + 1) \\
    & \times \Big\{J_0\Big(\frac{2A}{\omega_d}\Big)\sigma_z + 2\sum_{n=1}^{\infty}\Big[J_{2n}\Big(\frac{2A}{\omega_d}\Big) \text{cos}(2n\omega_d t) \sigma_z + J_{2n-1}\Big(\frac{2A}{\omega_d}\Big) \text{sin}([2n-1]\omega_d t) \sigma_y\Big]\Big\}.
\end{split}
\end{equation}
This expansion is exact and valid for any choice of parameters. We now do a rotating-wave approximation (RWA) by assuming that $\omega_d = \omega_m$ and $\omega_m \gg g$. After the first assumption there are 4 terms in $\tilde{H}$ that are not rotating, while the rest of the terms rotate with a frequency at least as high as $\omega_m$. The second assumption makes sure that the rotating terms average out to a small value on the timescale set by $g$ and can thus be ignored. By defining $\bar{A} \equiv 2A/\omega_d$ we recover Eq.~(3) in the main text.

\section{Logical states in the Fock basis}
In this section we show the Fock basis expansion of the logical states from Eq.~(5) and the calculation underlying Fig.~3 in the main text. The single mode squeezed vacuum state can be expanded in the Fock basis as 
\begin{equation}
    S(\xi)\ket{0} = \frac{1}{\sqrt{\text{cosh} r}} \sum_{n=0}^{\infty} (-1)^n \frac{\sqrt{(2n)!}}{2^n n! } e^{in \varphi} \text{tanh}^n r \ket{2n},
\end{equation}
where $\xi = re^{i\varphi}$ is the squeezing parameter. The logical states can then be written as 
\begin{align}
    \ket{0}_L &= \frac{1}{\sqrt{\mathcal{N}}_+}[S(\xi) + S(-\xi)]\ket{0}= \frac{2}{\sqrt{\mathcal{N}_+ \text{cosh}r}} \sum_{n=0}^{\infty} \frac{\sqrt{(4n)!}}{2^{2n} (2n)!} (\text{tanh}r)^{2n}  e^{i2n\varphi} \ket{4n},\\
    \ket{1}_L &=\frac{1}{\sqrt{\mathcal{N}}_-}[S(\xi) - S(-\xi)]\ket{0} = \frac{-2}{\sqrt{\mathcal{N}_- \text{cosh}r}}\sum_{n=0}^{\infty} \frac{\sqrt{(4n+2)!}}{2^{2n+1} (2n+1)!} (\text{tanh}r)^{2n+1}  e^{i(2n+1)\varphi} \ket{4n+2}.
\end{align}
The expectation value of $(b^\dagger b)^p$ in the two logical states become 
\begin{align}
    _L\bra{0}(b^\dagger b)^p \ket{0}_L &= \frac{4}{\mathcal{N}_+ \text{cosh}r} \sum_{n=0}^{\infty}\frac{(4n)!}{[2^{2n} (2n)!]^2} (\text{tanh} r)^{4n} (4n)^p,\\
    _L\bra{1}(b^\dagger b)^p \ket{1}_L &= \frac{4}{\mathcal{N}_- \text{cosh}r} \sum_{n=0}^{\infty}\frac{(4n+2)!}{[2^{2n+1} (2n+1)!]^2} (\text{tanh} r)^{4n+2} (4n+2)^p.
\end{align}
It is the ratio $_L\bra{0}(b^\dagger b)^p \ket{0}_L/ _L\bra{1}(b^\dagger b)^p \ket{1}_L$ that is plotted as a function of $r$ in Fig.~3 for $p\in \{1, 2, 3, 4\}$.

\section{Open system dynamics}
We provide here additional details on the open system dynamics and Fig.~4 in the main text. In principle, we assume the primary sources of decoherence to be excitation loss and gain, as well as dephasing of both the qubit and oscillator. However, we assume oscillator dephasing to be slow relative to the other decoherence processes, and it will therefore be ignored in the present context. Additionally, we ignore thermal excitation of the qubit as we consider an experimental setup where the qubit thermal occupation is small ($n_{q, th} \approx 10^{-7}$). The time evolution of the density matrix $\rho$ in the interaction frame is then described by the master equation 
\begin{equation}
    \frac{\partial \rho}{\partial t} = - i[H_1, \rho] + \frac{\gamma_\phi}{2}\mathcal{D}[\sigma_z]\rho + \gamma_1 \mathcal{D}[\sigma_-]\rho + \gamma_m n_{m,th} \mathcal{D}[b^\dagger] \rho + \gamma_m (n_{m,th} + 1) \mathcal{D}[b] \rho, 
\end{equation}
where $\gamma_1$ ($\gamma_m$) is the qubit (mechanical) damping rate, $\gamma_\phi$ is the qubit dephasing rate, and $n_{m,th}$ is the thermal occupation of the oscillator bath. $H_1$ is the interaction frame Hamiltonian given in Eq.~(\ref{eq:hamiltonian_interaction_frame}) and $\mathcal{D}$ is the Lindblad superoperator
\begin{equation}
    \mathcal{D}[\hat{O}] \rho = \hat{O} \rho \hat{O}^\dagger - \frac{1}{2}O^\dagger  O\rho  - \frac{1}{2} \rho O^\dagger O.
\end{equation}
We consider the system parameters given in the main text: $\omega_q = 20$ GHz, $\omega_m = 1$ GHz, $g = 100$ kHz, $n_{m,th} \approx 1$, $\gamma_m = 1$ kHz, and $\gamma_1, \gamma_\phi \in \{10, 100 \}$ kHz. We assume the drive to be optimal, i.e. $\omega_d=\omega_m$ and $2A/\omega_d = 2.405$.

We solve the master equation numerically up to a time $gt = 1.2$. The end time correspond to a squeezing strength $|\xi| \approx 1$ in the closed system. Fig.~4 (a) shows the fidelity between the even superposition prepared in an open vs. closed system for the four different combinations of $\gamma_1,\gamma_\phi$, and we see that the loss in fidelity is primarily associated with the transition from a superposition to a mixed state as illustrated with the Wigner plots in Fig.~4 (b)-(c).